\documentclass{aa}

\usepackage{times}
\usepackage{graphicx}
\usepackage{amsmath}
\def\teff{{T}_{\rm eff}}
\def\amlt{\alpha_{\rm MLT}}
\def\mast{{M}_\ast}

\def\msol{{M}_\odot}
\def\Mbol{M_{\rm bol}}
\def\lsol{{L}_\odot}
\def\zsol{{Z}_\odot}

\def\rast{{R}_\ast}
\def\last{{L}_\ast}
\def\mast{{M}_\ast}

\def\llsol{{L/L}_\odot}

\def\diff{{\rm d}}
\begin{document}

   \thesaurus{08     
              (
	       05.3;  
               12.3;  
               15.1;  
	       16.4   
              )} 
   \title{An Attempt to Pin Down the Instability Domain of Long-Period
          Variables}

\titlerunning{Instability Domain of LPVs}


   \author{Alfred Gautschy
          }


   \institute{Astronomisches Institut der Universit\"at Basel,
              Venusstrasse 7, CH-4102 Binningen
              \thanks{email: gautschy@astro.unibas.ch}
             }

   \date{Received ...... ; accepted .......}

   \maketitle

   \begin{abstract}

    The period-luminosity relation of Miras and semiregular variables
    in the Large Magellanic Cloud and in the Galaxy is used to locate
    their instability domain in the Hertzsprung-Russell diagram. We
    take advantage of the considerable width in luminosity of the
    relation at given period to assign masses to the observed
    long-period variables using stellar models on the the asymptotic
    giant branch and nonadiabatic pulsation computations. We study the
    sensitivity to chemical abundance of the position of the
    instability region on the Hertzsprung-Russell diagram. The mass
    function of the long-period variables along the AGB is discussed
    for Galactic and LMC variables.  Finally, we contribute to the
    dispute on the pulsation mode of Miras and lend support to the
    view that, for most of the Mira variables, the pulsation in the
    fundamental mode is more likely.


   \end{abstract}
%
%
\section{Introduction}

The first Mira variable was observed more than 400 years ago,
nevertheless our astrophysical understanding of the Miras as a class
is still far from satisfactory.  The lack of understanding pervades
stellar-evolution as well as pulsation aspects. Since these AGB
stars are very cool and extended, their envelopes are dominated by
convective energy transport and any successful theory explaining their
variability needs to incorporate a trustworthy pulsation-convection
interaction. To test the predictions of linear stability analyses
which incorporate some sort of convection interaction formalism
(e.g. Xiong et al. \cite{xdch98}), the results have to be compared
with the observed boundaries of the Mira instability domain on the
Hertzsprung-Russell (HR) diagram. Such an instability domain is
missing in the literature, probably partly due to the complexity of
assigning reliable temperatures to these large-amplitude variables at
low temperature.

Besides Mira variables with large-amplitude and long-pe\-ri\-od
pulsations, the AGB hosts~--~among others~--~also semiregular
variables. The members of subclass \textit{a} of the latter family
(SRa) look very much like Miras, except for the smaller
light-variability amplitudes.  In the following, we will collectively
refer to long-period variables (LPVs) when addressing Miras and SRas.

Astronomers agree on LPVs to be radial pulsators which oscillate
mostly in one mode only. But already the assessment of the radial
order of the pulsation mode in particular of Miras led to a
long-lasting and ongoing controversy in the literature. The
transformation of photometric color indices to temperatures is
difficult at the low temperatures of the Miras and the adopted choice
influences the derived pulsation mode.  Interferometrically determined
diameters and a low-temperature scale to assess surface temperatures
favor first overtone modes (e.g. Tuthill et al. \cite{tutetal94},
Haniff et al. \cite{hanetal95}, Feast \cite{feast96}, Barth\`es
\cite{barth98} and references therein). On the other hand, based on
period ratios between low- and high-amplitude long-period variables in
the LMC, Wood \& Sebo (\cite{ws96}) conclude that funda\-men\-tal-mode
pulsation is to be preferred for Miras.  For a few years now, also
theoretical studies favor Miras to pulsate in the fundamental mode
(e.g. Ya'ari \& Tuchman \cite{yt96}, Tuchman \cite{t99}). Today's
numerical treatment of convection still hampers the reliability of
these long-term simulations. But even along the linear-stability
avenue, the lack of a trustworthy pul\-sa\-tion-con\-vect\-ion
interaction prescription prevented robust predictions.  Only recently,
Xiong et al. (\cite{xdch98}) reported non-adiabatic results with a
sophisticated pul\-sa\-tion-con\-vect\-ion interaction
pre\-script\-ion. According to their results, the fundamental mode is
clearly preferred.

The period-luminosity (PL) correlation of LPVs on the AGB is
observationally well documented (e.g. Glass \& Lloyd Evans
\cite{gle81}, Feast et al. \cite{fgwc89} and the references therein,
Hughes \& Wood \cite{hw90}).  Currently, the true extension of the the
PL correlation at fixed period seems to be an unsettled issue. Feast
et al. (\cite{fgwc89}) presented a rather narrow relation whereas the
Hughes \& Wood (\cite{hw90}) data suggested a rather broad
band. Various observational aspects~--~insecure periods, random-phase
observations, biased data sets~--~can be responsible for the
discrepancy.  In any case, since LPVs are very luminous, they appear
as attractive distance indicators.  To exploit this potential, it
appears seducing to find filter passbands or phases during the
pulsation cycle which collapse whatever luminosity spread onto the
narrowest possible band (e.g. Feast \cite{feast84}, Hughes \& Wood
\cite{hw90}, or Kanbur et al. \cite{khc97}).  On the other hand, to
study stellar physics, we are going to take advantage of the
full spread of the PL relation and consider it as an additional source
of information.

Since Mira variables are thermally pulsing AGB stars they follow the
much referred to monotonous core-mass~--~luminosity relation only
during a short fraction of a thermal-pulse cycle (cf. Wagenhuber \&
Tuchman \cite{wt96}). On first sight, a simple parameterization of the
Miras' position on the AGB appears to be no longer possible due to the
non-monotonicity induced by the thermal pulses. On the PL plane,
however, the situation improves considerably.  During its numerous
successive thermal pulses, an AGB star traces out a locus on the PL
plane which almost collapses onto a single line. This line agrees very
well with one traced out by a star which increases its luminosity
monotonously as a function of core mass (see Fig.~3 of Wagenhuber \&
Tuchman \cite{wt96}). In other words, the locus traced out by an AGB
star on the PL plane is a unique function of mass, independent of the
phase during the thermal-pulse cycle.  We take advantage of this
property to assign masses to observed data and eventually map these
data back onto the HR plane; this is the programme of the following
paper.  In section 2, we describe the observations which we utilized,
the computational tools, and the assumptions entering the mapping of
LMC and Galactic Mira data onto the HR plane. Section~3 contains the
results which are discussed in Sect.~4. The latter section contains
also the presentation of the mass functions derived for oxygen- and
carbon-rich Miras in the LMC. The paper closes with some final
comments (Sect.~5) on the consequences of our determination of the
LPVs' instability region.

\section{Exploiting the PL relation}

To infer the position of the instability domain of LPVs on the HR
plane, we relied on luminosity estimates and period determinations for
LMC and Galactic variables. The LMC data were collected from Hughes \&
Wood (\cite{hw90}), referred to as HW90. From Feast et
al. (\cite{fgwc89}), abbreviated as FGWC89, we adopted a sample of LMC
Mira variables. The latter brightness estimates are phase
averaged. The HW90 magnitudes, on the other hand, were measured at
random phase; as the number of objects in this sample is considerably
larger, it is attractive to probe the extent of the instability
domain. The HW90 data contain not only large-amplitude Miras but also
SRa variables. Comparing the population of the PL plane by the two
families of pulsators reveals no systematic difference. Therefore, we
will not further distinguish between Miras and SRa's but we will treat
them equal and just refer to LPVs. It is worthwhile to stress that the
two data sets, the HW90 and FGWC89 one, complement each other in the
sense that the FGWC89 material tells something about the effect of
random-phase observations (as in HW90) onto the PL relation.  Despite
the about 50\% larger spread $\delta\!\log P$ at fixed $\log\llsol$ of
the HW90 data compared with the FGWC89 data set, we found no
systematic shifts between the two populations on the PL plane. The
origin of the considerably different $\delta\!\log P$ spread in HW90
and FGWC89 is an aspect which should be settled in the future as it
influences the evolutionary interpretation of the LPVs.

Galactic PL data are very scarce and therefore extensive analyses are
not possible to date. Nevertheless, in this paper we used the recently
HIPPARCOS-calibrated Galactic Mira data published by van Leeuwen et
al. (\cite{lfwy97}), referred to as vL97, and the results from
Robertson \& Feast (\cite{rf81}) (RF81).

\begin{figure}
      \resizebox{\hsize}{!}{\includegraphics{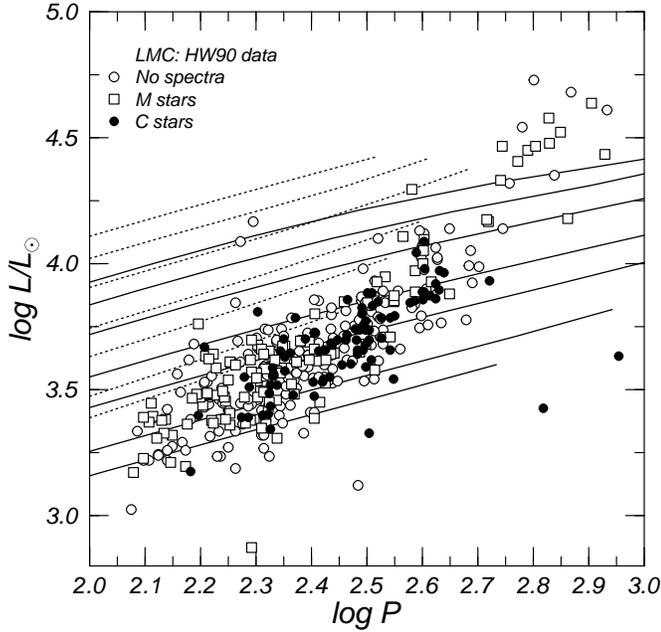}}
      \caption{PL data of HW90 observations and results from radial 
	       pulsation computations. F-mode data for
	       each mass choice is plotted as a solid line. The lowest
	       line results from the $0.8 \msol$ models. The masses
	       $1.0, 1.5, 2.0, 3.0, 4.0,$ and $5.0 \msol$ follow towards
               higher luminosities. The same ordering applies to
               O1 results displayed with dashed lines.  
              }
      \label{pl_hw90}
\end{figure}

To fit the observed PL data to theoretical data we performed
pulsation modeling as described below. The stars' loci along the
AGB relied on their parameterization by Vassiliadis \& Wood (\cite{vw93}):
\begin{equation}
  \begin{split}
\log {L / \lsol} = - 5 \log \teff & + 1.17 \log {M / \msol}  \\
		                      & - 0.581 \log {Z / \zsol} + 20.748 .
  \end{split}
\end{equation}
The set of masses on which the analyses are based comprise $0.8, 1.0,
1.5, 2.0, 3.0, 4.0,$ and $5.0 \msol$. For each mass, 20 models were
computed according to eq.~1 in the range of $3.8 < \log \teff <
3.4$. Envelope integrations were carried out to a temperature of
$10^6$ K. The surface boundary conditions were given by an atmosphere
integration with a pseudo-spherical T-$\tau$ relation. Convection was
treated in a non-adi\-aba\-tic MLT fashion (Baker \& Temeshvary
\cite{bt66}) and turbulent pressure was accounted for in the momentum
equation. The mixing-length was usually taken to be 1.5 pressure scale
heights ($H_{\rm P}$). The influence of other choices are discussed in
the text (Sect.~3). To set the magnitude of the isotropic turbulent
pressure we chose $A_\varpi = 1/3$ in our computations. The quantity
$A_\varpi \equiv p_{\rm turb} / \rho w_{\rm c}^2$ measures the
efficiency of generating turbulent pressure ($p_{\rm turb}$) out of
the convection-induced Reynolds stress ($w_{\rm c}$ stands for the
typical speed of the convecting eddies). Tests with different choices
of $A_\varpi$ revealed no significant changes of the tracks on the
$\Pi$L plane. In the following, we designate the theoretical
period-luminosity results with $\Pi$L to distinguish them from
observed PL data.  Noticeable effect of adding turbulent pressure to
the structure equations were observed only above about $\log L/\lsol =
3.75$.  At fixed luminosity and for masses above $3\msol$, the periods
of models computed with $A_\varpi = 0.5$ were shifted up by at most
about $0.07$ dex compared with model stars for which turbulent
pressure was neglected. For lower-mass models the shifts were
negligible (i.e. of the order of the thickness of the lines in
Figs.~\ref{pl_hw90} or \ref{pl_fgwc89}).

Opacity tables were taken from the OPAL 1992 release (Iglesias et al.
\cite{irw92}) which were extended with Alexander opacities at
low temperatures (Alexander \& Fergusson \cite{af94}). For the Galaxy
we assumed $Z = 0.02$ and $Z= 0.01$ for the LMC.

Linear non-adiabatic (LNA) radial pulsation computations were
performed with a Riccati-type code which derives from the version
described in Gautschy \& Glatzel (\cite{gg90}). For this study we
considered fundamental (F) and first-overtone (O1) modes. The
corresponding low eigenfrequencies still admit the use of reflective
boundary conditions at the surface to be a decent approximation.  This
was inspected by computing the radial pulsation cavity in the
adiabatic approximation.  The eigenfrequency calculations neglected
any pulsation-convection interaction.  This crude approach is
evidently inappropriate concerning imaginary parts of the
eigenfrequencies. In other words, with this approach we can neither
expect to determine the instability domain of Miras nor can we
contribute anything to the solution of the pulsation-mode
riddle. However, we are positive that the real parts of the
eigenfrequencies, i.e. the periods, are sufficiently accurate for our
purposes (cf. Xiong et al. \cite{xdch98}).

To derive the HR data and the mass function of the observed Miras on
the PL plane we inverted the data from the previously described LNA
analyses.  To compute birational spline mass surfaces (Sp\"ath
\cite{spaeth}) on the $\Pi$L plane we needed the data on a regular
rectangular grid. Therefore, we interpolated, with the
piecewise-monotonic cubics algorithm of Steffen (\cite{steffen}),
periods and luminosities at appropriate positions along lines of
constant mass.  For the stiffness parameter of the birational spline
we chose, after numerous experiments, always unity.  The resulting
mass, together with the observed period and an assumed abundance
(based on the stellar system to which the LPV belongs) allowed the
computation of an effective temperature according to the adopted AGB
parameterization. Such an effective temperature has to be interpreted
as the `equilibrium' value and it can differ systematically from some
phase-averaged \emph{mean} effective temperature over an observed
pulsation cycle. This aspect is further discussed in Sect.~4.

\section{Results}

First, we present the instability domains on the HR plane resulting
from the inversion of observed pulsation data mapped onto the $\Pi$L
plane. We start with the LMC and turn then to the modest Galactic
data. In this section, we concentrate mainly on F-mode inversions to
deduce the instability domain from the observations. If not noted
otherwise, we used a mixing-length of $1.5 H_{\rm P}$. 

\begin{figure}
      \resizebox{\hsize}{!}{\includegraphics{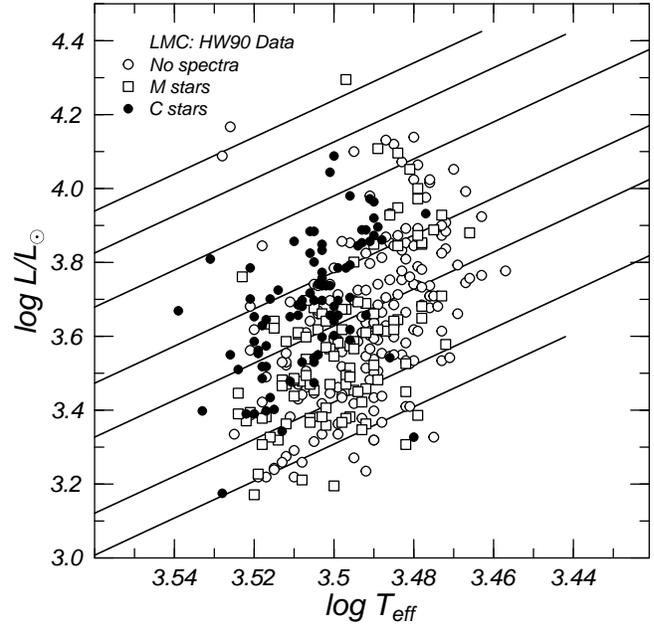}}
      \caption{The inverted instability domain traced out by the
	       HW90 data. The meaning of the symbols is the same
	       as in Fig.~\ref{pl_hw90}. The solid lines indicate
	       the loci of pseudo-evolutionary tracks on the HR plane
	       based on eq.~1. The lowest mass gives the lowest line
	       and they are ordered with increasing mass.
              }
      \label{is_hw90}
\end{figure}

Figure~\ref{pl_hw90} shows the PL (and $\Pi$L) diagram for the largest
available data set, the HW90 data. Observations are plotted with open
and filled symbols. The open ones represent either O-rich stars (also
referred to as M-type stars in the figures) or stars for which no
spectra (also abbreviated as NoSp stars) are available. C-rich
variables are plotted with filled circles. Full lines show
non\-adia\-batic radial F-mode periods as a function of luminosity for
a fixed stellar mass and for an O-rich composition appropriate for
$Z=0.01$.  Dashed lines represent the same information but for the
radial first-overtone (O1) mode. The various lines (of the same style)
are ordered with respect to the mass of the stellar model.  The lowest
mass ($0.8 \msol$) constitutes the lowest-luminosity line.  For the
inversion we restricted the period range to below about $500$ days and
the luminosities to below $32\,000 \lsol$. At longer periods the
simple-minded modeling of the stellar surface layers might no longer
be justified by the large mass-loss rates observed.  Within our
restricted domain of inversion we could perform the analyses in the
same fashion for the F as well as for the O1 modes.

\begin{figure}
      \resizebox{\hsize}{!}{\includegraphics{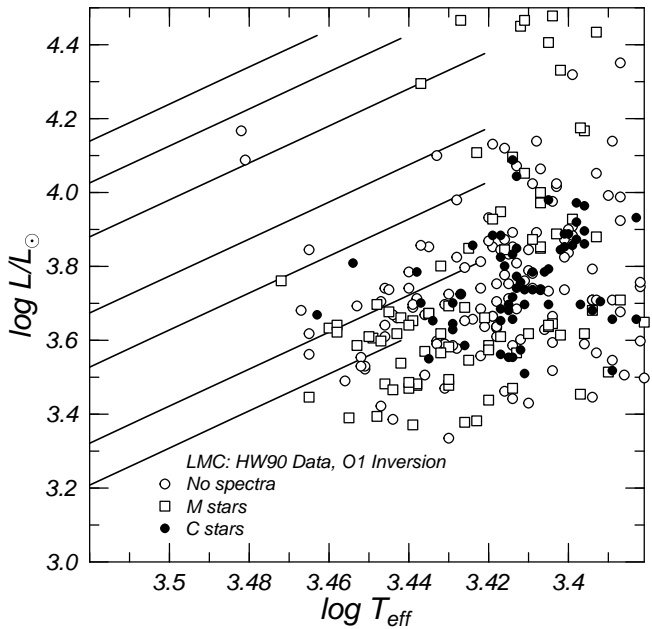}}
      \caption{Same as Fig.~\ref{is_hw90} but data was inverted with
	       the assumption of O1 pulsations for all
	       variables.
              }
      \label{is_hw90_o1}
\end{figure}

Figure~{\ref{is_hw90} shows the inverted observational data projected
onto the HR plane. We assumed all stars to pulsate in their F
mode. For the O-rich variables and those without spectral information
we used models with O-rich composition.  The solid lines correspond to
the tracks of different masses up the AGB, parameterized according to
eq.~1. The masses are the same as those mentioned in the caption of
Fig.~\ref{pl_hw90}. The lowest mass constitutes the lowest line and
the order is monotonic.  The C-rich stars were inverted with model
stars which were computed with an abundance ratio (by number) C/O $=
1.4$ but lying otherwise on the same tracks on the AGB as the O-rich
stars. As cautioned by Hughes \& Wood (\cite{hw90}), we found that the
C-rich variables are attributed different masses and effective
temperature than the O-rich ones when the different chemical
compositions are incorporated in the model stars.  This is
particularly expressed in the F mode; inversions based on O1 modes are
much less influenced by the composition choice. In case of the F mode,
the $\Pi$L line is decremented by at most $0.05$ dex at fixed period
at low masses and about half of that value above $4 \msol$.  If C-rich
variables were analyzed with O-rich models, somewhat lower masses and
lower effective temperatures were derived compared with an analysis
referring to appropriate C-rich models.

The blue border of the instability domain
is nearly vertical at $\log \teff \approx 3.53$. The red border, on
the other hand, might be skew, broadening the instability domain
towards higher luminosities. In any case, the red edge lies between
$\log\teff = 3.47$ and $3.45$.  As we have no information on the
completeness of the data sample we cannot estimate how robust the
latter statement is.

\begin{figure}
      \resizebox{\hsize}{!}{\includegraphics{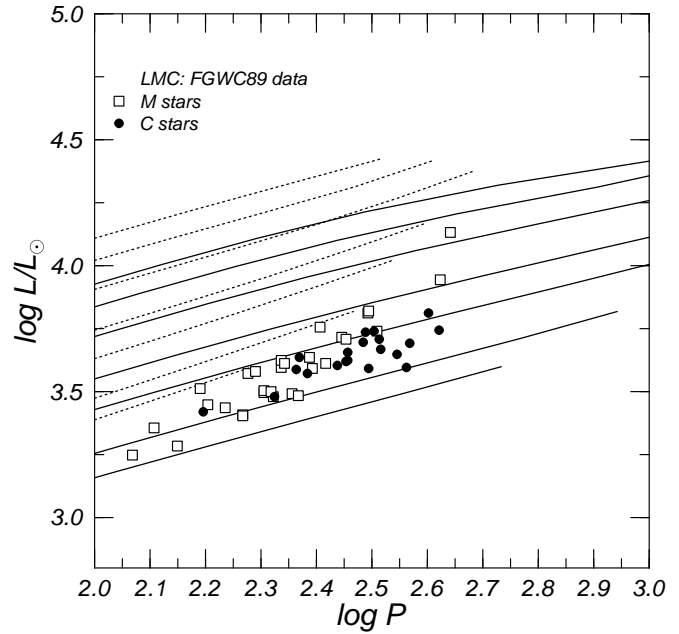}}
      \caption{Same as Fig.~\ref{pl_hw90} but with the
	       FGWC89 data. 
              }
      \label{pl_fgwc89}
\end{figure}

Figure~\ref{is_hw90_o1} shows the results of the inversion of the the
HW90 data assuming O1 modes only.  The emerging instability domain has
its blue boundary at $\log \teff \approx 3.47$. The boundary is not as
clearly defined as in Fig.~\ref{is_hw90}. A number of low-luminosity
data points had to be removed as they were located too far from the
lowest-mass $\Pi$L line so that the inversion failed.  The red edge is
located around $\log \teff = 3.38$.  Compared with Fig.~\ref{is_hw90},
the data points are much more scattered at low temperatures in
Fig.~\ref{is_hw90_o1}.  Even if this visual impression alone is a weak
argument against using the O1 assumption to find the instability
domain, we will further substantiate the use of the F mode in
Sects.~3.3 and~4.

The HW90 data are mostly based on single or few-epoch observations of
the corresponding long-period variable stars. The large pulsation
amplitudes of these variables might induce a vertical uncertainty in
the distribution of Miras in the PL relation and a corresponding
blurring of the true domain on the HR plane if only random-phase
observations are used. Fortunately, we can check for such an effect in
the HW90 data: Feast et al. (\cite{fgwc89}) published bolometric
luminosities of a significantly smaller set of LMC Miras whose
bolometric magnitudes were derived from averaging observations around
the whole light-cycle. Figure~\ref{pl_fgwc89} shows the corresponding
PL data superposed on the same pulsation computations that were used
for the HW90 data. Again, C-rich variables are plotted with a separate
symbol (filled circles) to distinguish them from O-rich variables.

\begin{figure}
      \resizebox{\hsize}{!}{\includegraphics{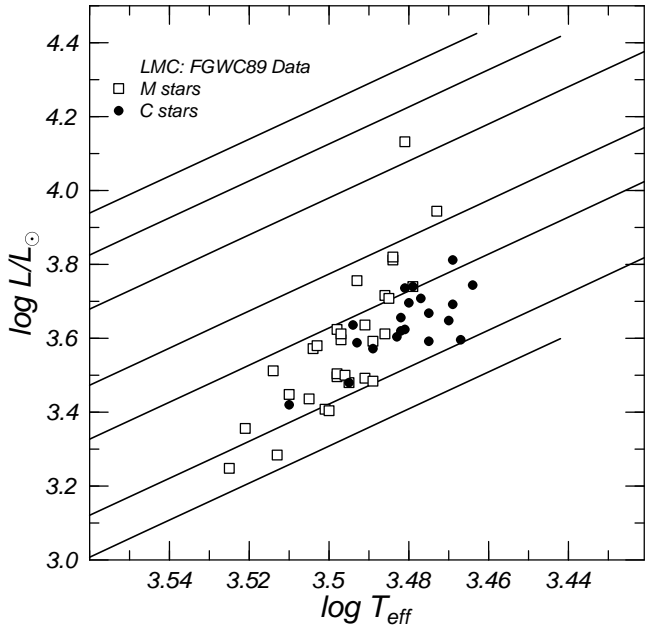}}
      \caption{Same as Fig.~\ref{is_hw90} but with the FGWC89
	       data.
              }
      \label{is_fgwc89}
\end{figure}

Figure~\ref{is_fgwc89} shows the sparsely populated HR domain after
inversion of the FGWC89 data with the F-mode pulsation assumption. In
contrast to HW90, the blue boundary~--~even if harder to
define~--~appears to be cooler towards higher luminosities.  A red
edge is hard to define; in any case, no variables were found below
$\log \teff \approx 3.46$. Despite the meager coverage of the PL
domain we can observe that the FGWC89 data lie comfortably in the
instability domain fenced by the HW90 data set.  In other words, we do
not see a systematic shift of the instability domains between the two
LMC data sets. The blue edge of the FGWC89 instability domain is,
however, clearly not vertical.

The analyses performed in the previous paragraphs are now applied to
Galactic data. However, reliable measurements of absolute magnitudes
are rare. In the following we discuss the Robertson \& Feast
(\cite{rf81}) and the recent HIPPARCOS-based measurements of $16$
galactic Miras published by van Leeuwen et al. (\cite{lfwy97}).  With
the available material we get at best a glimpse at the extent of the
instability domain. At this stage, we infer again the instability
domain from the assumption of F-mode pulsation only.

Despite the partial overlap only of the vL97 and the RF81 data-sets,
we superpose them in Fig~\ref{is_gal}. This improves at
least the appearance of the instability domain, even if its
reliability does not.

 \begin{figure}
      \resizebox{\hsize}{!}{\includegraphics{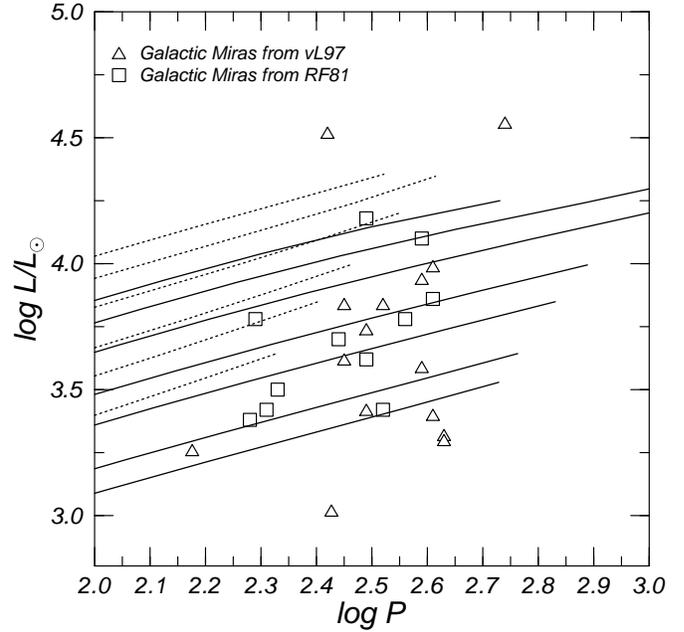}}
      \caption{Same as Fig.~\ref{pl_hw90}
	       for the Galactic data of RF81 and vL97. 
              }
      \label{pl_gal}
\end{figure}

\begin{figure}
      \resizebox{\hsize}{!}{\includegraphics{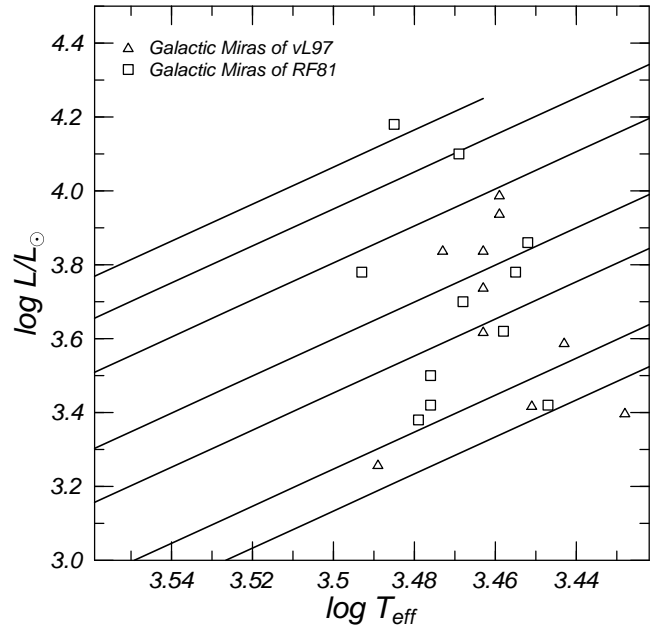}}
      \caption{ Same as Fig.~\ref{is_hw90} but with the Galactic
	       data of RF81 and vL97.
              }
      \label{is_gal}
\end{figure}

In spite of the low number of data points in each set we observe in
Fig.~\ref{is_gal} no systematic shift in $\teff$ between the two data
sources. This is of course a direct consequence of compatible
luminosity estimates~--~regardless of significant deviations of the
magnitude estimates for some objects in common in both sets
(cf. Fig.~\ref{pl_gal}). The Galactic Mira data shown in
Fig.~\ref{pl_gal} hint~--~again with the assumption of prevailing
fundamental modes~--~at a blue boundary of the instability domain at
$\log\teff \approx 3.49$. The red boundary is, once more, only ill
defined; it is certainly as cool as $\log \teff = 3.43$.

\subsection{The effect of thermally modified stellar models}
Recently, Ya'ari \& Tuchman (\cite{yt96}) argued~--~based on long-term
nonlinear model simulations~--~that Mira variables undergo mode
switching from O1 to F mode as late as a few hundred years after the
onset of the pulsational instability due to some nonlinear processes
of the (thermo-)dynamical action of the pulsation on the thermal
structure of the star. The mode switching is induced by the change of
stellar structure which is also reflected in the change of the
mean, as well as of the equilibrium radius of the star. Ya'ari \&
Tuchman conclude therefore that standard equilibrium stellar models,
usually used to compute Mira pulsation properties are inadequate. In
this subsection we investigate how strongly our inversion results
change subject to a systematic shift of a variable star's
equilibrium radius.

According to the results published by Ya'ari \& Tuchman (\cite{yt96})
the equilibrium radius of a model star lies close to the maximum
compression phase of the dynamical pulsation. By comparing initial data
with the state arrived at after about $500$ years we deduced that the
equilibrium radius shrinked by about $29$\%.  We modeled such a
shrinkage by modifying eq.~(1) such that the tracks of a given mass on
the HR plane were shifted to $29$\% smaller radii. This procedure leads
to the following relation:
\begin{equation}
  \begin{split}
\log {L / \lsol} = - 5 \log \teff & + 1.17 \log {M / \msol}  \\
		                      & - 0.581 \log {Z / \zsol} + 20.821 .
  \end{split}
\end{equation}
Compared with eq.(1), it is just the constant
term which changes upon modifying the radius at a given
luminosity. Models generated according to eq.~(2) will be referred to
as thermally modified (TM) hereafter.

Ya'ari \& Tuchman's (1996) result of pulsationally deflated radii
founded on a modified entropy structure of the stars' envelopes. Our
models have, at a given luminosity, a smaller radius than what is
predicted by canonical stellar evolution.  Otherwise their envelopes
are in thermal equilibrium. As we are interested in pulsation periods
only, i.e. not in stability properties, we are confident that this
crude approach is still a viable one.

\begin{figure}
      \resizebox{\hsize}{!}{\includegraphics{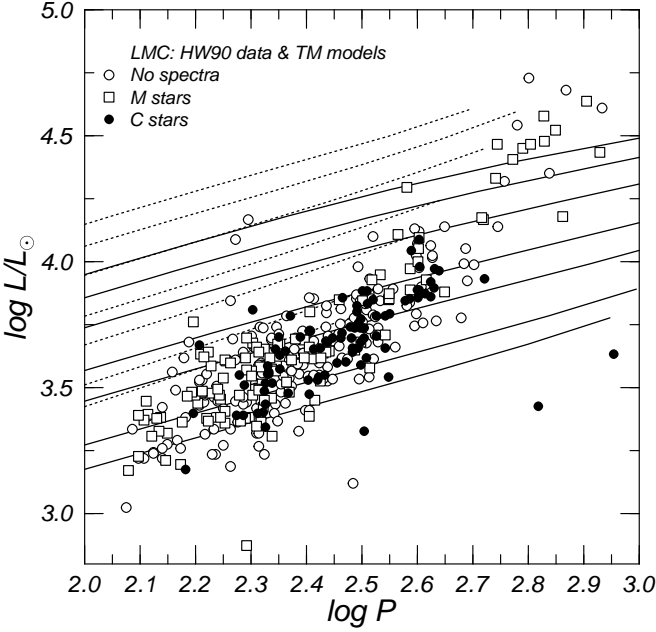}}
      \caption{ Same as Fig.~\ref{pl_hw90} but with thermally
	       modified stellar models.
              }
      \label{pl_hw90_tm}
\end{figure}

Computing the $\Pi$L relation for TM models revealed, comparing with
canonical models, only minor shifts of the observationally well
populated regions.  The F-mode lines (solid lines in
Fig.~\ref{pl_hw90_tm}) are parallel-shifted to higher luminosities by
at most $0.03$ dex at periods below about $300$ days.  For longer
periods the TM models show a steeper slope. For our inversion, this
domain was not crucial as we restricted ourselves to LPVs with periods
below $500$ days and the long-period population is rather small. As we
see in Fig.~\ref{pl_hw90_tm}, at fixed period the O1 lines are
shifted vertically with respect to the fiducial lines shown in
Fig.~\ref{pl_hw90} by about $0.05$ dex for all masses and for the
whole period range considered.

Mapping the HW90 data onto the HR plane using TM models
revealed~--~in accordance with the statements in the last
paragraph~--~that the locus of the instability domain which was
computed by inverting F-mode $\Pi$L data was hardly affected. The
blue border of the instability domain stayed at about $\log \teff =
3.52$. Only the red border shifted to slightly lower effective
temperatures, being now at around $\log \teff = 3.45$.

We will further defend the inversion with the F-mode assumption in
Sect.~4.  Be it sufficient here to say that when using standard model
stars in the analyses (as used in Sect.~3.1), the resulting preference
of F modes for LPV pulsations there implies an even stronger preference
of F modes in the case of using TM models.  This applies in
particular for the bulk of low-luminosity LPVs.

Notice that not only a deflation of a pulsating stellar envelope is
encountered in nonlinear simulations: Dorfi (\cite{ead98}) presented
preliminary results from radiation-hydrodynamical computations of a
luminous blue variable model whose equilibrium radius increased by
more than $10$\% during the first few dozen pulsation cycles. The
effect of an inflated equilibrium radius (compared with fiducial
evolutionary tracks) is studied in the following subsection; the
physical motivation to get there is a different one; the effect,
however, is the same.

\subsection{Changing the mixing-length}
Tuchman (\cite{t99}) argued in his review that pulsation results of
Miras, in particular the $\Pi$L relation, reacted sensitively on the
choice of convection modeling~--~i.e. it reacts on the choice of the
mixing length.  We agree with that statement if we force the model
star to remain at the same position on the HR plane during an MLT
modification.  However, it is not a priori obvious that the star's
position is not affected when its convection zones' mixing length is
modified.  Therefore, we performed a set of stellar evolution
computations with different mixing-length choices. Based on the
resulting tracks we re-parameterized their AGB loci in accordance with
the prescriptions used in eqs.~(1) and (2).  Tracks based on different
mixing lengths for the convection zones were parallel-shifted relative
to each other along the AGB. Models with larger mixing lengths
ascended the AGB at higher effective temperatures. We arrived at a
slope of $\Delta \log \teff / \Delta \alpha_{\rm MLT} = 0.056$ in the
range $1< \alpha_{\rm MLT} < 2$. The results from the evolution
computations and their transformation into the AGB parameterization
led again to a change of the constant term in eq.(1).  To study the
influence of a low mixing length ($\alpha_{\rm MLT} = 1$) on the
$\Pi$L relation we used the following AGB prescription:
\begin{equation}
  \begin{split}
\log {L / \lsol} = - 5 \log \teff & + 1.17 \log {M / \msol}  \\
		                      & - 0.581 \log {Z / \zsol} + 20.72 .
  \end{split}
\end{equation}

Figure~\ref{pl_hw90_amlt} shows that the LNA F-mode periods resulting
from the \mbox{$\alpha_{\rm MLT} = 1$} AGB sequences are considerably
longer at given luminosity and mass than what we deduced from,
e.g., Fig.~\ref{pl_hw90}. This applied to all masses considered. The O1
periods, on the other hand, dropped slightly compared with the
`fiducial' choice of $\alpha_{\rm MLT} = 1.5$ (and shown in
Fig.~\ref{pl_hw90}). In other words, reducing the mixing length
consistently, i.e. allowing also for modified evolutionary tracks, led
to an enlarged period separation between F and O1 modes.

\begin{figure}
      \resizebox{\hsize}{!}{\includegraphics{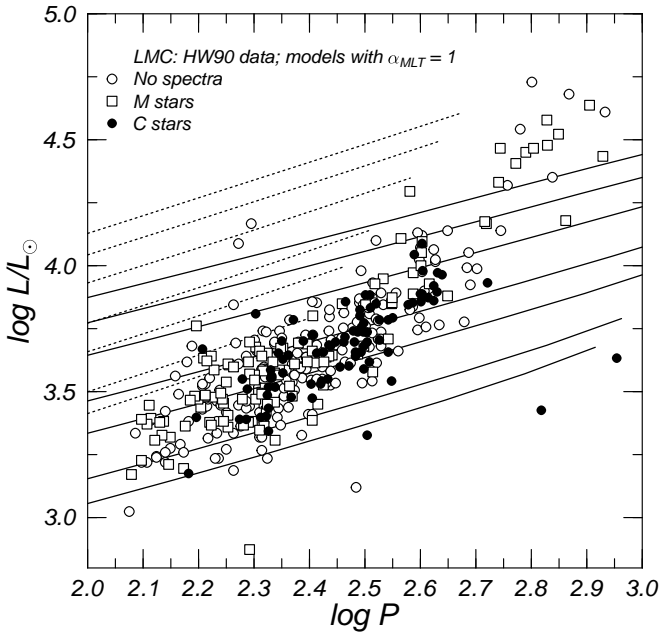}}
      \caption{ Same as Fig.~\ref{pl_hw90} but with stellar models
               having a convection zones parameterized with
	       $\amlt = 1$.
              }
      \label{pl_hw90_amlt}
\end{figure}

Hence, increasing the period at a chosen luminosity and fixed stellar mass
favored again the interpretation of short-pe\-ri\-od LPVs as F-mode
pulsators. On the $\Pi$L plane, the O1-period lines were shifted
considerably upwards and led to very low masses (below $0.5 \msol$) in
the mass-fit of the inversion procedure.  Therefore, we focus again on
the F-mode results to estimate the location and shape of the
instability domain.  The considerable shift of the $\Pi$L lines in
Fig.~\ref{pl_hw90_amlt} of the F modes translated into a shift of the
instability domain on the HR plane. The shift was purely horizontal
since the luminosities are fixed observationally.  We found the
$\alpha_{\rm MLT} = 1$ models to define an instability region which
was shifted to higher temperatures by $0.025$ in $\log\teff$ when
compared with the $\alpha_{\rm MLT} = 1.5$ (cf. Fig.~\ref{is_hw90}).
The shape of the region remained unchanged.

\subsection{The mass function of LPVs}
During the inversion procedure leading to the HR location of the data
on the PL plane we assigned masses to the observed variable stars.
Hence, we were in the position to derive a mass function of these
variables along the AGB.  As an example, we show the mass functions
that were obtained during the inversion of the HW90 data, once with
the assumption of F-mode pulsation (cf. Fig.~\ref{hw90_mfunf}) and
once with the assumption of O1 pulsation
(cf. Fig.\ref{hw90_mfuno1}). The two figures are representative for
all the the other data sets with which we worked.  Independent of the
choice of the observational data set, the shape of the distribution
functions remained rather stable.  The comparison of the HW90 data set
with the FGWC89 one revealed that only the maxima of the distributions
shifted slightly. This is rather surprising as we have to assume that
the various observational data sets suffer from different selection
effects.  It is indeed only the change of the assumption on the
pulsation mode which altered substantially the shape of the
distribution function.

The figures showing the mass distributions are subdivided into C-rich
stars and a plot collecting the rest of the sample, namely the O-rich
ones and the variables lacking detailed spectral classification
(NoSp).

For both inversions~--~assuming once F and then O1 modes~--~we
explored the mass function in the range between $0.5$ and $5
\msol$. For the F-mode pulsator assumption this domain embraces the
resulting mass range well. For the O1 case, on the other hand, the
number of low-mass pulsators increases rapidly and it continues so
also below $0.5$ $\msol$. As the inversion becomes numerically
problematic and physically questionable in this domain we ignored this
range.

\begin{figure}
      \resizebox{\hsize}{!}{\includegraphics{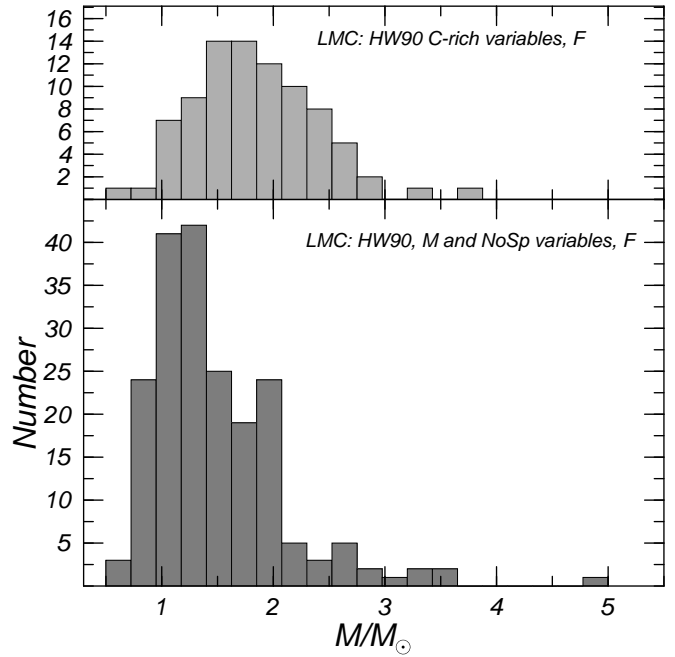}}
      \caption{Mass function along the AGB derived for the HW90
	       data set with the assumption of F-mode
	       pulsation only. The top panel shows the mass
      	       distribution for the C-rich stars in the sample.
	       The bottom panel is made up of all other stars
	       in the HW90 data (O-rich and stars without spectra).
               NoSp stands for no spectra available. 
              }
      \label{hw90_mfunf}
\end{figure}

Figure~\ref{hw90_mfunf} shows the well pronounced maxima of the mass
function for the C-rich as well as in the non-C variables. The C-rich
stars peak between $1.5$ and $1.8 \msol$ whereas the maximum for non-C
variables occurs at about $1.25 \msol$. Both distribution functions
show an asymmetric bell curve. The low-mass flank is steeper
than the high-mass one in both cases. Up to slight shifts of the
maxima, the results are surprisingly stable with respect to analyzing
different data sets. The FGWC89 data show the maximum of non-C
variables at $1.1 \msol$ and the one of the C-rich ones at about $1.4
\msol$.

When we performed the F-mode inversion with thermally modified
envelopes the character of the mass functions re\-main\-ed unchanged; only
the maxima shifted. Using the HW90 data on the TM models led to
maxima at about $1.3 \msol$ for the non-C \emph{and}
C-rich variables.  Performing the analysis on $\alpha_{\rm MLT} = 1$
models led to an upward shift of the maxima to $1.75 \msol$
for both the C-rich and the non-C variables, without any other
consequences for the distribution function.

For the Galactic data the mass functions are very uncertain due to the
low number of stars in the samples. We obtained crude results only for
the RF81 data, hinting at a mass-peak at about $1.25 \msol$.

\begin{figure}
      \resizebox{\hsize}{!}{\includegraphics{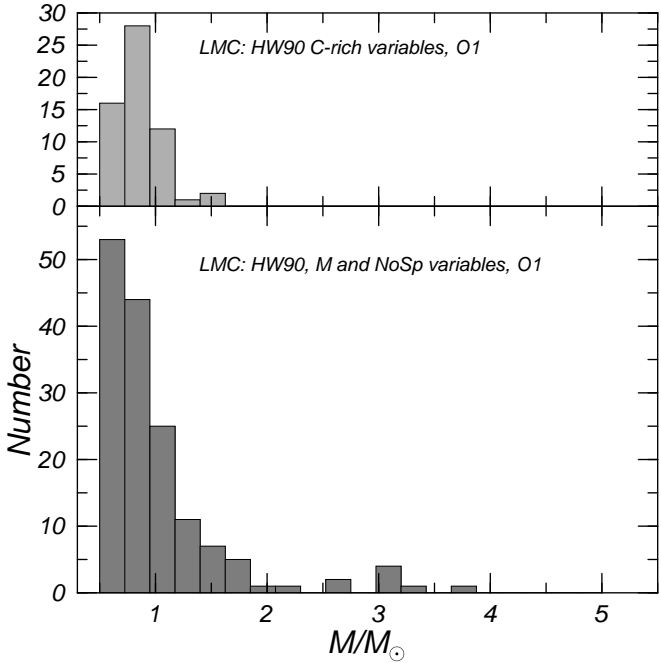}}
      \caption{ Same as Fig.~\ref{hw90_mfunf} but with the assumption
	       of O1 pulsations only.
              }
      \label{hw90_mfuno1}
\end{figure}

Despite the surprising agreement between the mass functions of
independently obtained data sets we must address their robustness with
respect to possible biases. This can be done only for the HW90 data
for which Hughes \& Wood (\cite{hw90}) investigated this aspect. So we
restrict our comments to this particular data set. Our mass function
must be compared with their `raw distributions' as a function of
period. Having to contemplate the raw data is not further restrictive
as it already reproduces the important aspects of showing a local
maximum and the shape of the short-period and long-period flanks of
the final distribution function which the authors (HW90) considered to
be representative for the LMC.  Assuming a unique relationship between
$\mast$ and period would let us easily transform their $\diff N /
\diff \log P$ plot (their Fig.~9) into our $N(\mast)$ plots
(cf. Figs.~\ref{hw90_mfunf} and \ref{hw90_mfuno1}). The Jacobian,
$\diff \log P / \diff \log \mast$, is of order of unity and does not
vary significantly over the period domain of interest so that the
shape of the number of variables as a function of period is preserved
along the mass coordinate. However, $\mast$ is not only a function of
period alone but it depends also on luminosity. Therefore, the
transformation is not that obvious and a preservation of the final
form of the mass function cannot be expected a priori. Nevertheless,
the number of long-period variable stars per mass interval seems to
retain the shape found in $\diff N / \diff \log P$.  The F-mode
inversion is further supported by comparing our $\diff N / \diff
\mast$ distribution with Iben's (\cite{iben81}) $\diff N / \diff
M_{\rm bol}$. The latter distribution shows also an asymmetric bell
shape which is preserved when transformed into a mass scale. The
Jacobian $\diff M_{\rm bol} / \diff \mast$ is of order unity and
varies only slowly with stellar mass.  Therefore, $\diff N / \diff
M_{\rm bol}$ and $\diff N / \diff \mast$ are functionally roughly
similar. If the aforementioned procedure is applied to the results of
the O1 inversion, $\diff N / \diff \Mbol$ would remain a purely
monotonously falling function towards brighter stars which does not
agree with the Iben (\cite{iben81}) results.

\section{Discussion}

In the previous sections we demonstrated how the instability domain of
LPVs can be deduced from period-luminosity data and linear
non-adiabatic pulsation computations along an AGB parameterization
which derives from stellar evolution calculations.  With this approach
we deduced effective temperatures which obtain for equilibrium
models. We analyzed data of different sources for Galactic and LMC
variables.

For the blue-edge position on the HR plane we claim to see a
metallicity dependence.  Reducing the heavy-element abundance shifts
the blue edge to higher effective temperatures.  The LMC blue edge is
about $200$~K hotter than the one for Galactic Miras.  The blueward
shift of the blue edge of the instability domain upon lowering the
heavy-element abundance is not attributable to the accompanying
blueward shift of the evolutionary tracks alone. At the position of
the blue boundary derived from LMC models we find evolutionary models
with Galactic chemical composition. The latter have higher mass,
however, and they do not yet pulsate. Only after further increase of
their luminosity, i.e., lowered their mean density in the envelope,
they start pulsating.

As an additional consistency test of our inversion we compared
temperatures of Galactic Miras~--~computed with F-mode and with O1
assumption~--~as a function of period with data presented in Feast
(\cite{feast96}). In this comparison we were interested in Galactic
Miras with temperatures deduced by means of angular diameter
measurements. These data present the `minimal'-assumption approach for
our purpose. The Galactic F-mode pulsators' temperature domain turned
out to be compatible with about half of the observed variables, the
rest of the observed stars was about $10$\% cooler than what the
inversion implied. When using the O1-assumption the situation
reversed: About half of the stars with angular diameters had low enough
effective temperatures to fit the inversion results and the rest
turned to be roughly $10$\% too hot.

We emphasize again that our effective temperatures correspond to
\emph{equilibrium} temperatures of the stars. These values can differ
substantially from the pulsation-cycle averaged mean temperature.
This point is emphasized in Fig.~\ref{sim_mira} where the nonlinear
pulsation cycle of a $1 \msol$ model star (model P7C14U4 of H\"ofner
et al. \cite{hjla98}) is shown. From the vanishing time-deri\-va\-tive
of the pulsational velocity we deduce the phases of passage through a
quasi-equilibrium state. These pha\-ses are indicated with arrows in
the bottom panel of Fig.~\ref{sim_mira}. The corresponding $\teff$
values are rather close to the maximum of the $\teff$ curve.
Therefore, as conjectured before, if simple averaging of temperature
measurements over a pulsation cycle is done, we expect the resulting
mean $\teff$ to be lower than the equilibrium values. In the model
star shown in Fig.~\ref{sim_mira} the difference amounts to about
$250$~K.

\begin{figure}
      \resizebox{\hsize}{!}{\includegraphics{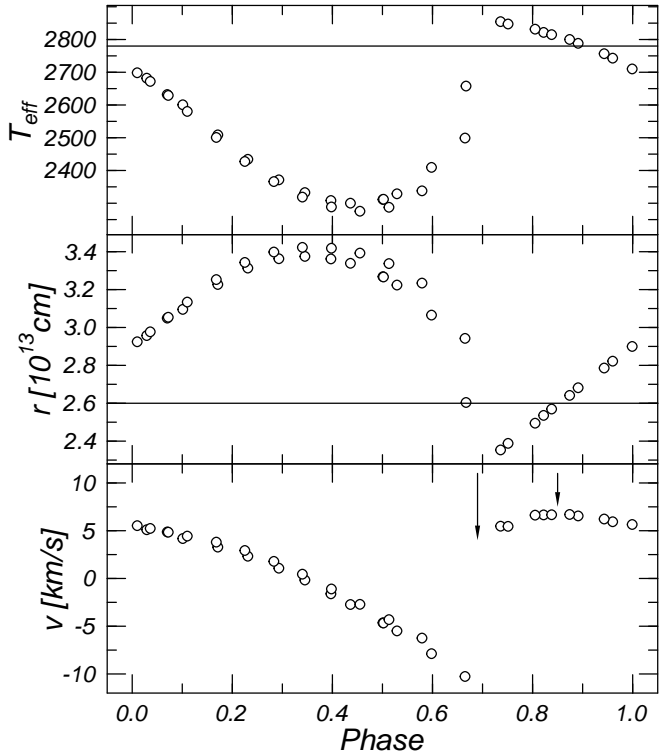}}
      \caption{Model data of a radiation-hydrodynamic simulation of a
      pulsating $1 \msol$ star at initially $7000 \lsol$. All physical
      variables at the photosphere are plotted over two periods which
      are folded onto the phase axis. The top panel shows the
      variation of the effective temperature. The middle panel
      displays the radius variation. The bottom panel contains the
      pulsational velocity of the photosphere. The two arrows in the
      bottom panel indicate the phases of passage through the
      equilibrium position of the model.  } \label{sim_mira}
\end{figure}

To finish the temperature discussion, we state our preference for the
temperature scale which is derived from the F-mode data. The
corresponding temperatures are slightly higher than what observations
indicate.  This, however, is in agreement with the discussion in the
last paragraph.  Notice that the slightly too high effective
temperatures (compared with the equilibrium values) which are derived
from observations are compatible with too large interferometrically
determined radii measured at random phase.  The middle panel of
Fig.~\ref{sim_mira} shows that the equilibrium radius is close to the
minimum radius and the mean~--~phase averaged~--~radius can exceed it
by at least $10$\%.

Along the process of deriving the instability domain we also computed
masses for observational data points. Hence, we assigned masses to LPV
stars and obtained mass functions for the different data sets. For a
given assumption about the pulsation mode of the variables we found
internally consistent functional forms of the mass functions for the
different LMC as well as for Galactic data sets. In both stellar
systems we arrived at a maximum of the mass functions at about $1.25
\msol$ for the O-rich variables. The mass distribution of C-rich
variables peaked between $1.5$ and $1.8 \msol$. All the distributions
resemble asymmetric bell-shaped functions with a steeper ascend on the
low-mass side and a more gradual decline on the high-mass flank.

Carbon stars were treated separately in the inversion process to
obtain their position on the HR plane.  Accounting explicitly for this
composition effect cau\-sed the observed C-rich variables to be
assigned higher masses and higher effective temperatures than what we
obtained when using O-rich models. The shift in mass, in particular
below about $3 \msol$, amounted to about $0.25 \msol$. The effective
temperature increased by about 0.025 in $\log \teff$. Treating the
C-rich variables separately cau\-sed them to concentrate in the hot
part of the instability domain. On the other hand, i.e. if treated the
same way as the O-rich variables, the C-variables tend to lie in the
cooler part of the instability domain.  In either case, we found the
C-variables to be slightly more massive than the O-rich LPVs.  The
functional forms of the mass function were comparable in both cases.
Our approach of treating the C-variables with chemically appropriately
modeled stars is only half way to the full solution as we still had to
rely on the canonical AGB-track parameterizations such as in
eq.~(1) which is based on O-rich stars.

\begin{figure}
      \resizebox{\hsize}{!}{\includegraphics{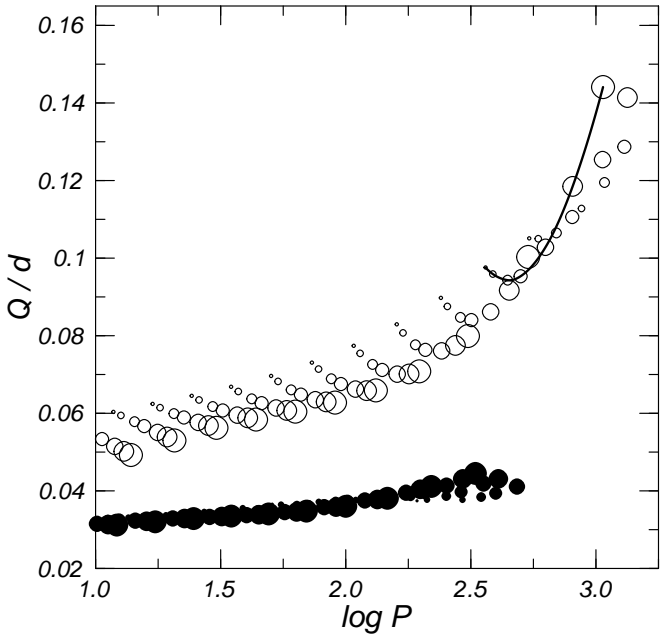}}
      \caption{Variation of the pulsation constant as obtained from
      the non\-adia\-batic pulsation computations for LMC composition,
      $A_{\varpi} = 0.5$, and $\alpha_{\rm MLT} = 1.5$. Open circles
      represent F modes and filled ones the O1 results. The size of
      the circles reflect the stellar mass. This plot is
      representative for all other computations.  } 
    \label{pq_lmc}
\end{figure}

Figure \ref{pq_lmc} shows the variation of the pulsation constant $Q$
as a function of the pulsation period for F (open circles) as well as
O1 (filled circles) modes of LMC models with $A_{\varpi} = 0.5$ and
$\alpha_{\rm MLT} = 1.5$. The size of the symbols characterizes the
model mass.  The F-mode pulsations define a band with a width of about
$0.02$ in $Q$ and it varies from roughly $0.05$ at $10$ days to $0.13$
at $1000$ days. The variation of the pulsation constant at fixed
effective temperature as a function of mass (and hence luminosity) is
traced out with the full line connecting the appropriate open
circles. We observe that the paraboloid is more expressed the cooler
the model stars are. In other words: In the low-luminosity range, the
pulsation constant drops upon increasing the star's mass. Once a
critical luminosity (or mass) is reached, the pulsation constant rises
again.  This behavior is also reflected in the normalized oscillation
eigenfrequency which shows a local maximum at the corresponding
mass. We can understand the phenomenon as follows: At low masses, the
mean-density decline upon increasing the stellar mass dominates over
the associated period increase. At higher masses (and large enough
luminosities) it is the pulsation-period lengthening with mass which
over-com\-pen\-sates the mean-density reduction. The rather rapid increase
of the pulsation period is attributable to the reduction of the mean
adiabatic exponent $\Gamma_1$ across the pulsation cavity due to the
growing contribution of radiation pressure which eventually forces the
period to infinity as the limit of $\Gamma_1 = 4/3$ is approached.

The O1 results in Fig.~\ref{pq_lmc}, plotted with filled circles, show
a much tighter correlation than the F-modes. The pulsation constant
$Q$ varies only between $0.03$ and $0.04$ in the period range from
$10$ to $300$ days. Additionally, the width in $Q$ at fixed period is
very small.  Functionally, however, the dependence of Q as function of
stellar mass at fixed $\teff$ is the same as for F-modes. Only the
amplitude of the variation is much smaller.

We conclude this section with a few elementary considerations of the
expected width of the PL relation of pulsators on the AGB.  Shibahashi
(\cite{shib93}) argued that the convergence of the stars' evolutionary
tracks making up the AGB induce a mass dependence and therefore a
substantial width into the PL relation. All the simple calculations
can be done referring to eq.~(1) and making use of the relation $\Pi
\propto \sqrt{\rast^3 /\mast}$.  We begin by assuming a perfect
coalescence of all evolutionary tracks on the AGB. In this case, any
instability strip appears as a line on the HR plane and it is
uniquely bordered by a lower and an upper luminosity. Even in this
case, the $\Pi$L~relation has a horizontal extension spanning
$\delta\!\log\Pi = -0.5\cdot\delta\!\log\mast$ and vertical one
extending over $\delta\!\log\last = 0.37\cdot\delta\!\log \mast$. The
horizontal extent is purely a consequence of the pulsations being
acoustic modes; the magnitude of the vertical width is influenced by
the steepness of the evolutionary tracks which is taken from
eq.~(1). In this framework, we expect the PL data to form a rhombus
with horizontal lower and upper border lines; the slope of the lateral
border-lines (i.e. $\diff\log\last/\diff\log\Pi$) is about 0.75.  When
we measure the horizontal spread in Figs.~\ref{pl_hw90} and
\ref{pl_fgwc89} we obtain a mass range $\delta\!\mast$ of about 0.8
(for HW90) and $0.55 $ (for FGWC89) assuming a typical mass of
$1\msol$ for these pulsating stars. These resulting mass ranges at
constant luminosity appear unrealistically high so that we consider
next a non-vanishing fan-out of the evolutionary tracks as a function
of mass.

At constant luminosity, $\Pi$ is now more sensitive to the star's mass
than in the first case. We find $\delta\!\log\Pi =
-1.16\cdot\delta\!\log\mast$. The mass range of variables at fixed
luminosity reduces now to the more reasonable values of $0.35 \msol$
for the HW90 and about $0.23 \msol$ for the FGWC89 data. The $\log
\last (\mast)$ dependence remains essentially the same as in the
coalescing-track scenario. In contrast to the simple first case, the
shape of the instability domain influences now the shape of the $\Pi$L
domain which is populated by LPVs. The narrower the instability strip
is, the smaller the luminosity spread at fixed period. Furthermore,
the larger the angle between the evolutionary tracks and the
borderline of the instability domain on the $\Pi$L plane, the stronger
the tendency for a large luminosity spread at fixed period even for a
narrow instability strip. Since the evolutionary tracks and the very
pulsational instability react on chemical composition, it is not
further surprising that noticeable differences of PL relations in
different stellar systems are reported in the literature (e.g. Menzies
\& Whitelock \cite{mw85}, Feast et al. \cite{fgwc89}). Adding up the
above aspects of widening the $\Pi$L relation of LPVs, 
the observation of a tight PL relation would be rather surprising.

\section{Conclusions}

We exploited the observed PL data by inverting them with the help of
theoretical $\Pi$L results and derived the instability domain on the
HR plane of LPVs in the LMC and of Miras in the Galaxy.

When inverting the PL data, model assumptions appeared to be not as
crucial as advocated in the past: In particular, we found no
significant effect of changing the mean radius of the model stars
which is expected to happen when they experience finite amplitude
pulsations (Ya'ari \& Tuchman \cite{yt96}). Furthermore, the choice of
the mixing length of convective eddies does not play an decisive
r\^ole~--~if it is applied consistently to \emph{evolutionary}
models. As expected, however, the choice of the pulsation mode
shifts the instability domain on the HR plane.  Assuming O1-modes to
be excited leads to an instability domain that is about $250$~K cooler
than the instability domain inferred from F-mode pulsations. The blue
edge of the latter instability region was found at about $\log \teff =
3.52$ for LMC abundances. Finally, also the choice of the chemical
composition is crucial.  Treating the C-rich variables separately from
the O-rich ones, turned out to be necessary. The position of the
C-rich pulsators shifted from the cool side in the instability domain
to the hot one when we performed the PL inversion with LNA results
distinguishing between C-rich and O-rich objects. The shape and the
location of the maximum of the mass function were, however, not
influenced significantly.  

As predicted by pulsation analyses including pulsation-con\-vect\-ion
interaction (Xiong et al. \cite{xdch98}) we encountered a shift of the
instability domain to higher effective temperatures upon reducing the
heavy-element abundances. In contrast to Xiong et al. (\cite{xdch98}),
we clearly locate a blue edge of the fundamental mode. 

The mass function which we obtained as a side-result peaks at about
$1.3 \msol$ for non-C variables and somewhere between $1.5$ and $1.8
\msol$ for C-rich variables if we imply F-mode pulsation. This result
is rather plausible. However, the numbers should be taken with
care. The mixed-mode interpretation mentioned in the last paragraph
and/or changing the chemical abundances will shift the maxima. In
particular, lowering the heavy-element abundances reduces the mass at
maximum. Nevertheless, the \emph{shape} of the distribution function
might be preserved.  In case of a mixture of pulsation modes
throughout the instability domain, we find O1 modes to be more likely
(regarding the inferred mass) in the upper luminosity range. 

The assumption that the mass function can be translated into a
luminosity function and the comparison of the result with the
literature led us to favor a fundamental-mode dominated explanation of
LPV pulsations.  A comparison of the predictions of Iben
(\cite{iben81}) with our results showed at least a structural agreement
of the F-mode mass function with the luminosity functions expected
from stellar evolution. Assuming a O1 dominated mass-function with its
exponential increase to lower masses appears unlikely according to
this exercise.

For Miras and SRas, the evolutionary scenario bringing them into their
instability domain remains obscure.  If the hypothesis that the LPVs
evolve in $\last$ and $P$ during their overstable phase applies
depends on the choice of the data set; the HW90 data (see
Fig.~{\ref{is_hw90}) would support it to some extend. In the FGWC89
this is much less pronounced, however.  In any case, for both data
sets, the instability domain is too extended in $\teff$ and $P$, when
measured along evolutionary paths, to justify the postulation of a
mass-dependent period at which the LPVs pulsate~--~without luminosity
evolution~--~before they leave the instability domain. Despite the
incompatibility with the Galactic and the LMC number density of
planetary nebulae (e.g. Wood \cite{w90}), but probably in better
agreement with the Miras' long-lifetime suggestion by Whitelock \&
Feast (\cite{wf93}), it appears necessary that Mira variables, or LPVs
in general, live through several thermal pulses. Otherwise, it is
unclear how they can populate the cool part of the instability domain.

The mass dependence in the LPVs' PL relation can be traced back to the
acute angle of the stars' evolutionary tracks along the AGB at which
they enter the instability domain and the mutual convergence of the
tracks along the AGB.  For a star's pulsation it is its total mass
which is relevant; at the AGB the mass cannot be parameterized as a
function of luminosity due to the near convergence of the evolutionary
tracks and therefore it cannot be replaced in the
period~--~mean-density relation. The radius appearing in the latter
equation can be substituted by a luminosity parameterization of a
characteristic locus in the instability domain (such as e.g. the blue
edge or a ridge line). Eventually, we arrive at a $P(\mast, \last)$
relation (cf. Shibahashi \cite{shib93}). 


\begin{acknowledgements}
      This project was supported by the Swiss National Science
      Foundation through a PROFIL2 fellowship. H. Harzenmoser
      contributed generously to the project with Gstaader B\"argch\"as
      surchoix and much appreciated deep\-er insights into the physics
      of stellar interiors.  Peter Wood kindly sublet the HW90 data in
      electronic form which speeded up the starting phase
      considerably.  Rita Lo \!\textsf{IDL}ed generously the Mira
      simulation data and critically commented on
      the manuscript. Constructive critique by M. Feast, K. Schenker,
      and W. L\"offler was helpful to bring the script to its final
      form.
\end{acknowledgements}



\end{document}